\documentclass{article}
\usepackage{spconf,amsmath,graphicx}
\usepackage{amsfonts}
\usepackage{amsmath}
\usepackage{subfigure}
\usepackage{mathrsfs}
\usepackage{multirow}

\usepackage{mathtools,lipsum,cuted}
\usepackage{xfrac}
\usepackage[colorlinks]{hyperref}
\usepackage[short]{optidef}
\usepackage{bbm}
\usepackage{tikz}
\usepackage{float}

\newsavebox{\tempbox}

\usepackage[]{caption} % Figures caption
\usepackage{graphicx}
\usepackage[export]{adjustbox}
\usepackage{caption}

\usepackage[ruled]{algorithm}
\usepackage{algpseudocode,algorithmicx}

\newcommand{\vz}{{\boldsymbol z}}
\newcommand{\vA}{{\boldsymbol A}}
\newcommand{\vR}{{\boldsymbol R}}
\newcommand{\vr}{{\boldsymbol r}}
\newcommand{\vP}{{\boldsymbol P}}

\newcommand{\vs}{{\boldsymbol s}}
\newcommand{\round}[1]{\ensuremath{\left \lfloor#1 \right \rceil}}

\title{Geometric Invariants for Sparse Unknown View Tomography}
\name{Mona Zehni, Shuai Huang, Ivan Dokmani\'{c}\thanks{This work is partially supported by National Science Foundation under Grant CIF-1817577.}, Zhizhen Zhao}
\address{Department of ECE and CSL, University of Illinois at Urbana-Champaign}

\begin{document}
\savebox{\tempbox}{\begin{tabular}{@{}r@{}l@{\space}}
&\scriptsize{SNR}\\ \scriptsize{$M$}
\end{tabular}}

\newcommand{\remove}[1]{}

\ninept
\maketitle

\begin{abstract}
In this paper, we study a 2D tomography problem for point source models with random unknown view angles. 
Rather than recovering the projection angles, we reconstruct the model through a set of rotation-invariant features that are estimated from the projection data. For a point source model, we show that these features reveal geometric information about the model such as the radial and pairwise distances. This establishes a connection between unknown view tomography and unassigned distance geometry problem (uDGP). We propose new methods to extract the distances and approximate the pairwise distance distribution of the underlying points. We then use the recovered distribution to estimate the locations of the points through constrained non-convex optimization. Our simulation results verify the robustness of our point source reconstruction pipeline to noise and error in the estimation of the features.

% Not having access to the projection angles alongside with the inherent sparsity of the model adds to the difficulty of this problem. For this purpose, we estimate a set of rotation invariant features from the projection data which no longer depend on the projection angles. We will show that these features are only a function of the geometry of the model, i.e. the radial and pairwise distances of the point sources. Next, we propose two methods that extract the distance information from the invariant features efficiently. Based on the recovered distances, the problem of locating the point sources is expressed as a constrained non-convex optimization which is then solved through projected gradient descent. We show the robustness of our distance extraction methods to various levels of noise. Furthermore, we empirically observe that as long as there is a minimum separation between the distances, the extraction of the geometry information is accurate. Also, we show the results of the full pipeline, verifying its functionality.
\end{abstract}

\section{Introduction}
\begin{figure*}[t!]
    \centering
    \includegraphics[width = 0.85 \linewidth]{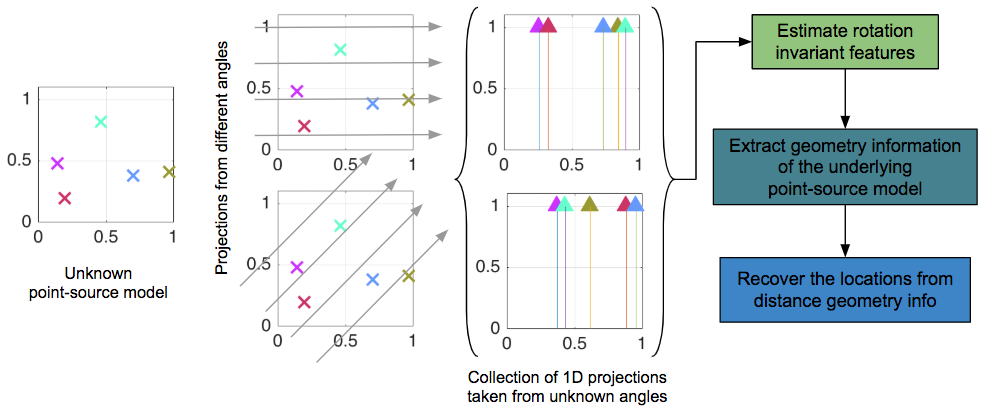}
    \vspace{-2mm}
    \caption{The 2D point source localization pipeline. Crosses indicate point sources; triangles indicate Dirac deltas in 1D.}
    \label{fig:pipeline}
    \vspace{-2mm}
\end{figure*}

We consider the following forward model,
\begin{align}
    s_\ell [u] &= \mathcal{D}  \{ \mathcal{P}_{\theta_\ell} I \}[u] + \varepsilon_\ell [u], \quad \ell \in \{1,2,...,L\} \nonumber \\
    I(x,y) &= \sum\limits_{k = 1}^{K} \delta(x-x_k,y-y_k)
\end{align}
where $I$ is %$I:\mathbb{R} \times \mathbb{R} \rightarrow \mathbb{R}$ denotes
an unknown model consisting of point sources (i.e. Dirac delta distributions) located at $\{(x_k,y_k)\}_{k=1}^{K} \in \mathbb{R}^2$, and $\mathcal{D}$ is the discretization operator. For the sake of simplicity, we assume unit weights for all the point sources. The operator $\mathcal{P}_{\theta}$ is the 1D projection along direction $\theta$, where $\theta$ is the angle between the projected direction and the horizontal $x$ axis. We assume that $\theta_\ell$ is a sample drawn from a uniform distribution over $[0,2\pi)$ which is a realistic assumption in many scenarios. %We introduce $\mathcal{D}: C_0(\mathbb{R}) \rightarrow \mathbb{R}^{M}$ to take into account of the finite resolution in digital imaging. It presents the sampling operator that maps from the space of continuous real-valued functions to a finite sequence of real values. Additionally,
We introduce the sampling operator $\mathcal{D}$ to take into account of the finite resolution of the digitized projection data, 
\begin{align}
\mathcal{D}(f)[u] = \int\limits_{\left(u - \frac{1}{2}\right)\Delta}^{\left(u + \frac{1}{2}\right)\Delta} f(x) dx, \, \text{for }u \in \{-M, \dots, M\},
\end{align}
where $\Delta$ is the sampling step-size.  The observed signals are contaminated by additive white Gaussian noise, i.e. $\varepsilon_{\ell}[u] \sim \mathcal{N}(0,\sigma^2)$. Fig.~\ref{fig:pipeline} illustrates the observation model.

% We consider the following forward model,
% \begin{align}
%     y_\ell [u] &= \mathcal{D} \left(\mathcal{F}  \{\mathcal{P}_{\theta_\ell} I\} \right) [u] + \varepsilon_\ell [u], \quad \ell \in \{1,2,...,L\} \nonumber \\
%     I(x,y) &= \sum\limits_{k = 1}^{K} \delta(x-x_k,y-y_k)
% \end{align}
% where $I:\mathbb{R} \times \mathbb{R} \rightarrow \mathbb{R}$ denotes the unknown point source model consisting of point sources (i.e. Dirac delta distributions) located at $\{(x_k,y_k)\}_{k=1}^{K}$. $\mathcal{P}_{\theta}$ operates as the 2D radon transform along $\theta$ direction and $\mathcal{D}: C_0(\mathbb{C}) \rightarrow \mathbb{C}^{M}$ denotes the sampling operator that maps from the space of continuous complex functions $C_0(\mathbb{C})$ to a finite sequence of complex values. Additionally, $\mathcal{D}(f)[u] = f(\Delta u)$ for $u \in \{0,1,...,M-1\}$ and $\Delta$ denoting the sampling step. Note that $\theta$ indicates the angle between the direction along which 2D radon transform is taken and the x axis of the xy-coordinate system. Here we assume that $\theta_\ell$ is a sample drawn from a uniform distribution over $[0,2\pi)$. Finally, the measurements are further contaminated by additive iid Gaussian noise, i.e. $\varepsilon_{\ell}[u] \sim \mathcal{N}(0,\sigma^2)$. Figure \ref{fig:pipeline}-(a) further illustrates the observation model.

Our goal is to recover the point source model from $\{s_\ell\}_{\ell=1}^L$. The problem of recovering a point-source signal arises in array signal processing \cite{Krim1996}, compressed sensing \cite{Boche2015}, super-resolution \cite{Candes2012}, radio astronomy \cite{Hanjie2017,Leap2017}, unassigned distance geometry \cite{Billinge2016,Duxbury2016}, imaging molecules using X-ray crystallography \cite{Drenth2007}, powder diffraction \cite{Fabian2006} and cryo-electron microscopy (EM) atomic modeling \cite{Scheres2012}, to name a few. %The main motivation behind our observation model comes from atomic model recovery in Cryo-EM \cite{Scheres2012}. In Cryo-EM the ultimate target is to recover the atomic model of a molecule from hundreds of thousands noisy projection images that are taken from random poses. 

%More specifically, consider the problem of recovering the sub-molecular model of a protein (i.e. protein folding) \cite{Dill2012}. Currently, achieving the folding of a protein from cryo-EM images involves an intermediate step of recovering the 3D density map from the projection images. Next, the protein sequence (amino-acid sequence) is docked into the density map such that it is compatible with molecular dynamic force fields \cite{Greevy2016}. Thus, one can imagine that our approach is a step toward direct recovery of the atomic model from projection data in a lower dimensional set-up (2D rather than 3D) and eliminates the intermediate step of recovering the density map.

The state of the art approaches in 2D tomography from known projection angles mainly recover the underlying image by solving a regularized least squares problem \cite{Donati2017,Nilchian2013}. The regularization then imposes the structure of the image, for examples its smoothness or sparsity. The main assumption of these methods is the availability of the projection angles. On the other hand, 2D tomography from unknown projection angles is addressed in \cite{Basu_Feas,Basu2000,Coifman2008} which aim at recovering the projection angles from the projection data and then invert the tomography operation to recover the image. However, these methods do not incorporate the model of the underlying image, for example, sparsity in our parametric image model. 
% Basu et al in \cite{Basu_Feas}, \cite{Basu2000} addresses the 2D tomography from projection data with unknown viewing angles. There are several differences between our approach and theirs, 1) our proposed method is specialized for point-source models, 2) they solve the tomography problem by recovering the viewing angles while we recover the parameters of our model directly, 3) our approach is parametric which targets the recovery of the parameters of the model rather than the whole underlying image ($I$). 

In this paper we propose a pipeline that recovers a 2D point source model directly from a set of projection lines taken at random unknown angles. 
Our method consists of three steps, 1) constructing rotation-invariant features from the projection data, 2) estimating distances and distance distributions from the rotation-invariant features, and 3) reconstruction of the point source model from the estimated distance distributions. Through the use of rotation-invariant features, we circumvent the recovery of the projection angles. We show that the estimated features are expressed as a summation of zeroth-order Bessel functions of the first kind. We then use the properties of Bessel functions to estimate the unlabeled radial/pairwise \textit{distances values} and their \textit{distance distributions}, which can be efficiently computed using the proposed methods. Specifically, the unlabeled radial distance distribution describes the distribution of the distances from the point sources to the origin, i.e. the center of mass. The unlabeled pairwise distance distribution, also known as pair distribution function (PDF), describes the probability distribution of the pairwise distances between any two point sources. In the final step of our pipeline, following the same line as \cite{Huang2018}, we recover the point source model from the estimated distance distributions by solving a constrained nonconvex optimization problem. The proposed pipeline is robust to noise if the number of samples are sufficiently large. 
% To circumvent the recovery of the projection angles corresponding to each projection, we construct a set of rotation invariant features from the projection data. From these features which are in the form of a summation of the scaled zeroth-order Bessel functions, it is possible to restore some geometry information of the model regarding the radial and pairwise distances or distance distributions. For this purpose, we suggest two computationally efficient methods which rely on the properties of the Bessel functions. Note that the set of pairwise distances recovered at this stage are unlabeled. The unlabeled pairwise distances, implicitly summarize as pair distribution function (PDF). PDF describes the probability distribution of the pairwise distances between the point sources. Next, locating the point sources from the recovered distances is formulated as a non-convex optimization problem, following the same lines as \cite{Huang2018}. 

\remove{
The organization of the paper is as follows. In section \ref{sec:method} we describe our pipeline. We present the results of our approach in section \ref{sec:results} and then conclude the paper in section \ref{sec:conclusion}. 
}

\section{Method}
 \label{sec:method}
We start by elaborating upon the generation of the invariant features from the projection data. Next, we describe two methods that target estimating the distances and the distribution of the distances. Finally, we describe the point source model reconstruction via the estimated distance distributions.
 
\subsection{Estimating the rotational invariant features}
We estimate a set of rotational invariant features from the observations $\{s_{\ell}\}_{\ell=1}^{L}$. These features are functions of certain geometric information of the point source model, namely, the radial distances from the points to the origin and the pairwise distances between any two points. We assume that the center of mass is at the origin and all pairwise distances are smaller than $R$. All points lie within a disk of radius $R$. Since the point source model $I$ is the summation of Dirac deltas, its projections from different angles also consist of Dirac deltas,
\begin{align}
    (\mathcal{P}_{\theta} I)(r) = \textstyle\sum\limits_{k=1}^{K} \delta(r - \left(  y_k \cos \theta - x_k \sin \theta \right)). \label{eq:proj_conti}
\end{align}
The observation is digitized into $2M + 1$ bins and the discretization operator $\mathcal{D}$ applied to \eqref{eq:proj_conti} leads to, 
\begin{align}
g_\theta[u] = & \left( \mathcal{D}  \mathcal{P}_{\theta} I \right)[u] = \textstyle\sum\limits_{k=1}^{K} \mathbbm{1}_{\frac{y_k \cos \theta - x_k \sin \theta}{\Delta} \in \left [ u - \frac{1}{2}, u + \frac{1}{2} \right ]}, 
\label{eq:final_obs}
\end{align}
where $u \in \{-M, \dots, M\}$ and $\Delta = \frac{2 R}{2M + 1}$. 
%\begin{align}
%& \left( \mathcal{D}  \mathcal{P}_{\theta} I \right)[u] = \sum\limits_{k=1}^{K} \int\limits_{u \Delta - \frac{\Delta}{2}}^{u \Delta + \frac{\Delta}{2}} \delta(r - \left(  y_k \cos \theta - x_k \sin \theta \right)) dr \nonumber \\
%& = \sum\limits_{k=1}^{K} \int\limits_{-\infty}^{\infty} \delta(r - \left(  y_k \cos \theta - x_k \sin \theta \right)) g(r-u \Delta) dr \nonumber \\
%& = \sum\limits_{k=1}^{K} g\left(( \frac{y_k \cos \theta - x_k \sin \theta}{\Delta} - u) \Delta \right) \nonumber \\
%& = \sum\limits_{k=1}^{K} \textbf{g} \left[\frac{y_k \cos \theta - x_k \sin \theta}{\Delta} - u \right] \label{eq:final_obs}
%\end{align}
%where $\textbf{g}[u] = g(u \Delta)$ for $u \in \{-M,...,M\}$ and $g(r) = 1$ if $r \in [-\frac{\Delta}{2},\frac{\Delta}{2}]$ and zero otherwise.
In order to derive the rotation-invariant features from \eqref{eq:final_obs}, we first take the discrete Fourier transform (DFT) of a projection line at angle $\theta$,
\begin{align}
\hat{g}_{\theta}[\nu] 
& = \textstyle\sum\limits_{k=1}^{K} \exp \left (\imath \frac{2 \pi \nu} {(2M+1)} \round{ \frac{y_k \cos \theta - x_k \sin \theta}{\Delta}} \right),
\label{eq:g}
\end{align}
where $\round{\cdot}$ denotes the nearest integer. For sufficiently small $\Delta$, we can approximate~\eqref{eq:g} without the rounding. 
Assuming the view angles are uniformly distributed, the rotation-invariant features are defined as:

\begin{align}
\mu[\nu] &= \mathbb{E}_{\theta} \{\hat{g}_{\theta}[\nu]\} \approx \textstyle\sum\limits_{k=1}^{K}  J_0 \left(\frac{ \pi r_k}{R} \nu \right )\,, \label{eq:final_mu_features}
% & = \sum\limits_{k=1}^{K} \textbf{G}[-\nu] \mathbb{E}_{\theta} \{\exp(j \frac{2 \pi}{2M+1} \frac{y_k \cos \theta - x_k \sin \theta}{\Delta} \nu) \} \nonumber \\
% & \stackrel{(a)}{=} \sum\limits_{k=1}^{K} \textbf{G}[-\nu] \frac{1}{2 \pi} \int\limits_{0}^{2 \pi} \exp(j \frac{2 \pi}{2M+1} \frac{y_k \cos \theta - x_k \sin \theta}{\Delta} \nu) d \theta \nonumber \\
% & \stackrel{(b)}{=} \sum\limits_{k=1}^{K} \textbf{G}[-\nu] J_0(\frac{2 \pi r_k}{\Delta (2M+1)} \nu) \label{eq:final_mu_features}
\end{align}
where $J_0(\cdot)$ is the zeroth order Bessel function of the first kind. The radial distance of the $k$-th point source is denoted by $r_k$, i.e., $r_k = \sqrt{x_k^2 + y_k^2}$. Following the same procedure, we derive the power spectral density as,
\begin{align}
 \mathbb{E}_{\theta} \{\vert \hat{g}_{\theta}[\nu] \vert^2\} & \approx \textstyle\sum\limits_{m=1}^{K} \textstyle\sum\limits_{n=1}^{K} J_0\left(\frac{ \pi d_{m,n}}{R} \nu \right) \nonumber \\
 & = K J_0(0) + 2 \textstyle\sum\limits_{m=1}^{K} \textstyle\sum\limits_{n=m + 1}^{K} J_0\left(\frac{ \pi d_{m,n}}{R} \nu \right),
\end{align}
where $d_{m,n}$ represents the distance between the $m$-th and $n$-th point source, $d_{m,n} = d_{n,m} = \sqrt{(x_n-x_m)^2+(y_n-y_m)^2}$. 

Therefore, we define the second-order invariant feature as,
\begin{equation}
    C[\nu] =  \left(\mathbb{E}_{\theta} \{\vert \hat{g}_{\theta}[\nu] \vert^2\} - K \right) /2\,.
    \label{eq:final_C_features}
\end{equation}
We estimate the features in \eqref{eq:final_mu_features} and \eqref{eq:final_C_features} from the observations $\{s_\ell[u]\}_{\ell=1}^{L}$ as,
%\begin{align}
%\widehat{\mu}[\nu] &= \frac{1}{L} \sum\limits_{\ell=1}^{L} {s}_\ell[\nu] \stackrel{L \rightarrow \infty}{\longrightarrow} \mu[\nu] \nonumber \\
%\widehat{C}[\nu] &= \frac{1}{L} \sum\limits_{\ell=1}^{L} \vert {s}_\ell[\nu] \vert^2 - (2M+1)\sigma^2 \stackrel{L \rightarrow \infty}{\longrightarrow} C[\nu]
%\label{eq:rot_inv_features}
%\end{align}
\begin{align}
&\widehat{\mu}[\nu] = \frac{1}{L} \textstyle\sum\limits_{\ell=1}^{L} \hat{s}_\ell[\nu],\\
&\widehat{C}[\nu] = \left(\frac{1}{L} \textstyle\sum\limits_{\ell=1}^{L} \vert \hat{s}_\ell[\nu] \vert^2 - (2M+1)\sigma^2 - K\right )/2\,,
\label{eq:rot_inv_features}
\end{align}
where $\hat{s}_\ell[\nu]$ is an empirical realization of $\hat{g}_\theta[\nu]$, and the subtraction of $(2M+1) \sigma^2$ in \eqref{eq:rot_inv_features} serves to debias the estimation of $C$. By the law of large numbers, the sample estimates $\hat{\mu}$ and $\hat{C}$ converge to $\mu$ and $C$ when the sample size $L \rightarrow \infty$. The features are invariant to the global rotation of the point sources.  %Note that the set of features presented in \eqref{eq:final_mu_features} and \eqref{eq:final_C_features} are rotation invariant meaning that if the point source model is globally rotated in the space then these features will not change. This fact is also revealed from the form of the features, as they are only functions of the radial and pairwise distance, which are invariant to global rotations.

\subsection{Recovering the geometry information of the model}
We propose two methods to extract the geometric information of the model from the invariant features. 

$\bullet$ \textit{Prony-based distance estimation (PBDE):} In our first method, we make use of the asymptotic behavior of the Bessel function \cite[p.364]{Abramowitz1972},
\begin{align}
    J_0(z) \approx \sqrt{\frac{2}{\pi z}} \cos \left(z - \frac{\pi}{4} \right),
    \label{eq:bessel_approx}
\end{align}
for $z \gg 1/4 $, to extract $\{r_k\}_{k=1}^K$ and $\{d_{m,n}\}_{m,n=1}^K$ using Prony's method.
%\frac{(1-i)e^{ix}+(1+i)e^{-i x}}{2 \sqrt{\pi x}}
Based on \eqref{eq:bessel_approx}, for sufficiently large integer $\nu$, the features can be approximated by,
\begin{align}
 \label{eq:feature_bessel_approx} 
    \widehat{\mu}[\nu]  \approx \textstyle\sum\limits_{k=1}^{K} \frac{e^{\imath \left(a_k \nu -\pi / 4 \right)} + e^{-\imath \left(a_k \nu - \pi/4\right)}}{\sqrt{2\pi  a_k \nu }},
\end{align}
where $a_k = \frac{\pi r_k}{R}$, for integer $\nu \gg \lceil \frac{1}{4a_k}\rceil$. We scale $\widehat{\mu}[\nu]$ by $\sqrt{\nu}$ and the new features $\sqrt{\nu} \widehat{\mu}[\nu]$ are approximated by the sum of $2K$ complex exponentials. Thus, we apply Prony's method to determine the filter that annihilates $\widehat{\mu}$. Arguments of the complex zeros of this filter correspond to the radial distances. %A summary of this procedure is presented in Algorithm \ref{alg:prony_based}. 
We can use the same procedure to recover the pairwise distances from $\widehat{C}$.
% recover the harmonics involved in the features. Next, we find the roots of the polynomial formed by the annihilating filter. Finally, based on the roots of the polynomial that are located on the unit circle, i.e. have unit norm, we extract $\{r_k\}_{k=1}^{K}$ and $\{d_{m,n}\}_{m,n=1}^K$.  

$\bullet$ \textit{Distance distribution estimator (DDE):} In our second method, rather than recovering the distances explicitly, we derive a distance distribution function that is later used to recover the locations of the point sources. For this purpose, we rely on the orthogonality properties of the Bessel functions,
\begin{align}
    \delta(u-v)= u \int_0^\infty tJ_0(u t)J_0(v t)dt, \, \forall{v} \in \mathbb{R}\,. \label{eq:bessel_property}
\end{align}
In order to implement \eqref{eq:bessel_property}, we replace the infinite upper limit of the integral with a frequency cutoff $c$, 
\begin{align}
 &\int_0^c tJ_0(u t)J_0(v t)dt = c \frac{u J_1(u c)J_0(v c) - v J_0(u c) J_1(v c)}{u^2 - v^2}  \,.
 \label{eq:bessel_fint}
\end{align}
The integral~\eqref{eq:bessel_fint} is evaluated using the Gauss-Legendre
quadrature rule~\cite[Chap. 4]{press1992numerical} and we perform Hankel transform on the features $\mu$ and $C$,
\begin{align}
 f_\mu (u_j) = \int_{0}^c \mu(t) J_0(u_j t) t dt \approx \sum_{i = 1}^n w(t_i) t_i \widehat{\mu}(t_i) J_0( u_j t_i), \nonumber \\
 f_C(u_j) = \int_{0}^c C(t) J_0(u_j t) t dt \approx \sum_{i = 1}^n w(t_i) t_i \widehat{C}(t_i) J_0( u_j t_i)\,, \label{eq:transformed_features}
\end{align}
 with $n$ points $\{t_i\}_{i = 1}^n$ on the interval $[0, c]$ and the associated weights $w(t_i)$. The Fourier coefficients at the non-equally spaced quadrature points are evaluated from the projection lines using the non-uniform FFT~\cite{greengard2004accelerating}. We define the approximated distributions of the radial distances and the pairwise distances in \eqref{eq:density_mC} as,
 \begin{align}
 p_\mu(u_j) = \frac{\vert f_\mu(u_j) \vert^2 }{  \sum_{j = 1}^l \vert f_\mu(u_j) \vert^2  }, \, p_C(u_j) = \frac{\vert f_C(u_j) \vert^2 }{  \sum_{j = 1}^l \vert f_C (u_j) \vert^2  }, \label{eq:density_mC} 
 \end{align}
 where $l$ is the total number of equally spaced points on the real line, and use $p_\mu$ and $p_C$ to recover the point source model in the next subsection.

\subsection{Point source recovery from distance distribution}
% \begin{figure}[tpb]
%     \centering
%     \includegraphics[width=0.4\columnwidth]{images/discrete_domain.pdf}
%     \caption{The 2D domain space $\mathbb{R}^2$ is divided into $m$ unit cells, each cell corresponds to a possible point location.}
%     \label{fig:discrete_domain}
% \end{figure}
After the radial and pairwise distance distributions are computed, we can recover the locations of the point sources using the approach proposed by \cite{Huang2018}. %As shown in Fig. \ref{fig:discrete_domain}, 
The 2D domain space $\mathbb{R}^2$ is first divided into $m$ unit cells $\{c_i\}_{i = 1}^m$. The point source locations are then represented by an indicator vector $\vz\in\{0,1\}^m$ where each cell corresponds to a possible point source location. Each entry $z_i$ of $\vz$ corresponds to a unit cell $c_i$ in the domain space $\mathbb{R}^2$; $z_i=0$ implies that no point source occupies the $c_i$ cell, while $z_i=1$ marks the existence of a point in the corresponding cell. 

Based on this definition, the pairwise distance distribution $Q(d)$ can be written in closed-form with respect to $\vz$:
\begin{align}
    Q(d)=\frac{1}{m^2}\vz^\mathrm{T}\vA_d\vz\,,
\end{align}
where $\vA_d\in\{0,1\}^{m\times m}$ is a symmetric Toeplitz matrix. Let $d_{c_i,c_j}$ denote the distance between the $i$-th cell $c_i$ and the $j$-th cell $c_j$. The $(i,j)$-th entry of $\vA_d$ is determined by
\begin{align}
    A_d(i,j)=\left\{
    \begin{array}{l}
    1, \\
    0,
    \end{array} \quad 
    \begin{array}{l}
    \textnormal{if }d_{c_i, c_j} = d\\
    \textnormal{if }d_{c_i, c_j}\neq d\,.
    \end{array}
    \right.
    \label{eq:A}
\end{align}
The pairwise distance distribution $p_C(d)$ is extracted from the projection lines using \eqref{eq:density_mC}. Now the target is to find an indicator $\vz$ such that its corresponding pairwise distance distribution $Q(d)$ matches $p_C(d)$. For this purpose, as proposed in \cite{Huang2018}, we relax the integer constraint on $\vz$ and minimize the nonconvex cross entropy between $p_C(d)$ and $Q(d)$ subject to a set of convex constraints,
\begin{align}
\begin{split}
    \min_\vz\quad&-\sum_dp_C(d)\log Q(d)\\
    \textnormal{subject to}\quad&\|\vz\|_1=n\textnormal{ and }z_i\in[0,1], \forall i\\
%    &0\leq x_i\leq 1,\, \forall i\\
    &\vR\vz= \vr \\
    &\vP \vz=\vs\,,
\end{split}
\label{eq:udgp_formulation}
\end{align}
where $\vR$ is a matrix such that $\vR\vz=\vr$ enforces the radial distance distribution $p_\mu(d)$, represented by the probability vector $\vr$; while $\vP$ denotes the projection operator that relates the point source model $\vz$ to one of the observed projection lines $\vs$. The matrix $\vR$ is derived similarly as~\eqref{eq:A}. We use projected gradient descent to solve~\eqref{eq:udgp_formulation}.
\begin{algorithm}[t!]
  \caption{Point source recovery from distance distribution} \label{alg:udgp} 
    \textbf{Input:} Projection data, $\{s_{\ell}[u]\}_{\ell=1}^{L}$ \\
    \textbf{Output:} The estimated $\{x_k, y_k\}_{k=1}^{K}$
  \begin{algorithmic}[1]
    \State Estimate the distance distributions, $p_{\mu}$ and $p_C$, from $\{s_{\ell}[u]\}_{\ell=1}^{L}$ using \eqref{eq:rot_inv_features} and DDE.
    \State Incorporate $p_{\mu}$ and $p_C$ to formulate \eqref{eq:udgp_formulation} and solve it using projected gradient descent.
 %   \State Report the recovered $\{r_k\}_{k=1}^{K}$ based on the roots located on the unit circle and identified by the Prony method.
  \end{algorithmic}
\end{algorithm}

\section{Numerical results}
\label{sec:results}
\begin{figure*}
    \centering
    \includegraphics[width = 0.88 \linewidth]{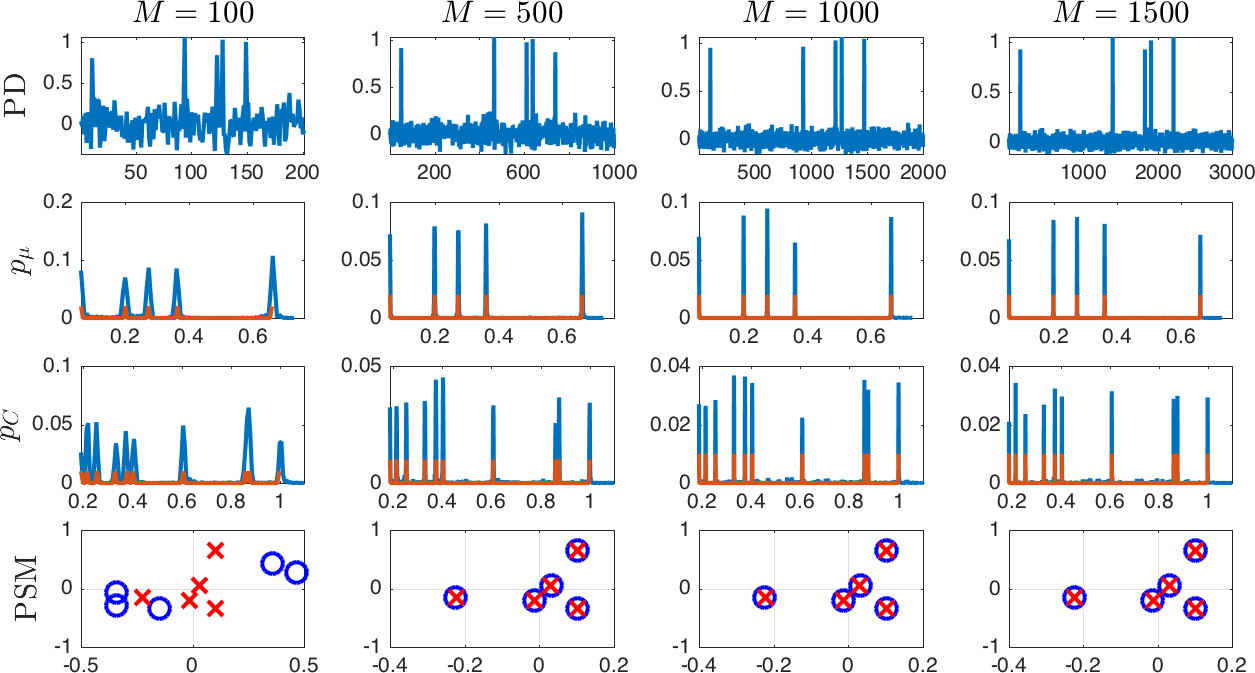}
    \caption{An illustration of the projection data (PD, the first row) for $\mathrm{SNR} = 1$, the estimated distribution of the radial and pairwise distances \eqref{eq:density_mC} (the second and third row), a comparison of the true and the recovered point source models (PSM, the last row). In the second and third row, the blue and orange curves represent the true and estimated distance distribution (up to a scaling factor). In the last row, the blue circles and red crosses mark the recovered and true point source models.}
    \vspace{-2mm}
    \label{fig:results_1}
\end{figure*}

\begin{table}
    \caption{Success rate (\%) of PBDE in recovering the radial distances}
    \begin{minipage}{.5\linewidth}
      {$K = 5$}
      \centering
\[\begin{array}{c|cccc}
\tikz[overlay]{\draw (0pt,\ht\tempbox) -- (\wd\tempbox,-\dp\tempbox);}%
\usebox{\tempbox}\hspace{\dimexpr 1pt-\tabcolsep}
& \infty & 100 & 10 & 1  \\
\hline
100 & 95 & 88 & 80 & 53 \\
500 & 99 & 90 & 76 & 50\\
1000 & 98 & 90 & 80 & 56\\
1500 & 98 & 89 & 77 & 52\\
\end{array}\]
    \end{minipage}%
    \begin{minipage}{.5\linewidth}
      \centering
    {$K = 10$}
\[\begin{array}{c|cccc}
\tikz[overlay]{\draw (0pt,\ht\tempbox) -- (\wd\tempbox,-\dp\tempbox);}%
\usebox{\tempbox}\hspace{\dimexpr 1pt-\tabcolsep}
& \infty & 100 & 10 & 1\\
\hline
100 & 49 & 47 & 43 & 36\\
500 & 65 & 53 & 48 & 38\\
1000 & 72 & 55 & 49 & 38\\
1500 & 76 & 52 & 48 & 38\\
\end{array}\]
    \end{minipage} 
    \label{tab:prony}
\end{table}

% remove the following block because the table won't fit
\remove{
\begin{table}
    \caption{Success rate (\%) of the point source recovery from distance distribution}
    \begin{minipage}{.55\linewidth}
      {$K = 5$}
      \centering
\[\begin{array}{c|ccccc}
\tikz[overlay]{\draw (0pt,\ht\tempbox) -- (\wd\tempbox,-\dp\tempbox);}%
\usebox{\tempbox}\hspace{\dimexpr 1pt-\tabcolsep}
& \infty & 100 & 10 & 1 & 0.5 \\
\hline
100 & 25 & 28 & 26 & 22 & 7\\
500 & 95 & 94 & 93 & 93 & 81\\
1000 & 97 & 96 & 96 & 94 & 80\\
1500 & 95 & 97 & 95 & 98 & 90\\
\end{array}\]
    \end{minipage}%
    \begin{minipage}{.5\linewidth}
      \centering
    {$K = 10$}
\[\begin{array}{c|ccccc}
\tikz[overlay]{\draw (0pt,\ht\tempbox) -- (\wd\tempbox,-\dp\tempbox);}%
\usebox{\tempbox}\hspace{\dimexpr 1pt-\tabcolsep}
& \infty & 100 & 10 & 1 & 0.5\\
\hline
100 & 0 & 0 & 0 & 0 & 0\\
500 & 75 & 73 & 73 & 73 & 47\\
1000 & 80 & 81 & 82 & 75 & 49\\
1500 & 92 & 92 & 91 & 90 & 58\\
\end{array}\]
\end{minipage} 
\label{tab:udgp_res}
\end{table}
}

\begin{table}
    \caption{Success rate (\%) of the point source recovery from distance distribution}
    \vspace{-5mm}
    \centering
    \[\begin{array}{c|ccccc|ccccc}
    \multicolumn{1}{c}{}    &\multicolumn{5}{c}{K=5} &\multicolumn{5}{c}{K=10}\\
    \tikz[overlay]{\draw (0pt,\ht\tempbox) -- (\wd\tempbox,-\dp\tempbox);}%
\usebox{\tempbox}\hspace{\dimexpr 1pt-\tabcolsep} 
    & \infty & 100 & 10 & 1 & 0.5 & \infty & 100 & 10 & 1 & 0.5\\ \hline
    100 & 25 & 28 & 26 & 22 & 7 & 0 & 0 & 0 & 0 & 0\\
    500 & 95 & 94 & 93 & 93 & 81 & 75 & 73 & 73 & 73 & 47\\
    1000 & 97 & 96 & 96 & 94 & 80 & 80 & 81 & 82 & 75 & 49\\
    1500 & 95 & 97 & 95 & 98 & 90 & 92 & 92 & 91 & 90 & 58\\
    \end{array}\]
\label{tab:udgp_res}
\end{table}

To test the algorithm performance, we generate the coordinates of $K$ points randomly on $\left[-1, 1 \right]\times \left[-1, 1\right] $. To generate the features, we take uniformly distributed $L= 10^4$ projections of the point source model. We generate $100$ random point source realizations. The invariant features are computed from \eqref{eq:rot_inv_features}, followed by \eqref{eq:transformed_features} and \eqref{eq:density_mC} in order to get the transformed features and an estimation of the distance distribution. Also, we define signal to noise ratio (SNR) of the projection data as the average power of the clean projection line divided by $\sigma^2$.

We use earth mover's distance (EMD) \cite{Rubner2000} to quantify the performance of our methods. EMD is a measure of the distance between two probability distributions. We say a recovery is successful if the EMD between the recovered distance distribution and the true one is smaller than a threshold $th=0.1$. Accordingly, we define success rate as the portion of trials for which successful recovery is achieved,
\begin{align}
    \textrm{success-rate} = P\{\textrm{EMD}(p,q) \leq th\}
\end{align}
where $p$ and $q$ mark the true and the recovered distributions.

\subsection{Discussion of the results}
There are two ways to extract geometric features from the random projections:

$\bullet$ \textit{Extracting the radial distances using PBDE:} Table \ref{tab:prony} present the success rate in the recovery of radial distances for $K = 5$ and $K = 10$ respectively for different values of SNR and $M$. In order for \eqref{eq:bessel_approx} to be a good approximation, we choose $\nu \geq 10$. For the Prony based method, $M$ denotes the number of discretizations of the projection lines. Thus, larger $M$ means finer discretization of the projection lines. The results suggest that as $K$ increases, it is harder to extract the radial distances due to denser distance distributions. On the other hand, as long as the approximation in \eqref{eq:bessel_approx} is accurate, changing $M$ does not significantly affect the performance. Furthermore, the higher the SNR, the higher the success rate. This happens as a result of more accurate estimation of the features from the projection data.

$\bullet$ \textit{The performance of the point source reconstruction:} For this experiment, the estimated features $\widehat{\mu}$ and $\widehat{C}$ are transformed by \eqref{eq:transformed_features}. Then we use $p_\mu$ and $p_C$ defined in~\eqref{eq:density_mC} and solve the optimization problem~\eqref{eq:udgp_formulation} using projected gradient descent to find the locations of the points. Fig.~\ref{fig:results_1} summarizes the observations, approximate distance distribution and the final results of our pipeline for a setting of $K = 5$ and $\mathrm{SNR} = 1$. Note that although the projection data (first row) are severely contaminated by noise, the features (second and third row) are estimated accurately and match features from the true point source model very well (up to a scaling factor). 
%In addition, effect of $M$ on the performance of the pipeline (failed recovery for $M = 100$). 
%Smaller $M$ means coarser discretization of the projection data which reduces the accuracy in estimating \eqref{eq:transformed_features}. 
In addition, coarser discretization of the projection data (smaller $M$) reduces the accuracy in extracting the geometric invariants and localizing the points  (see the first column of Fig.~\ref{fig:results_1}).

Table \ref{tab:udgp_res} shows the success rates in the point source recovery of the model from distance distributions.  
The results suggest that 1) it is critical to have a large $M$ in order to have a good approximation of the transformed features in \eqref{eq:density_mC}, 2) our pipeline is robust to various noise levels on the projection data, 3) with larger $K$ the success rate decreases, because the minimum separation between points become smaller.

\section{Conclusion}
\label{sec:conclusion}

In this paper, we proposed a pipeline to recover a point source model from a set of projections taken from unknown angles. Instead of first recovering the angles and then solving a tomography problem, we directly recover the locations of the point sources using a set of rotational invariant features that are estimated from the projection data. The features contain geometric information about the point source model: the unlabeled radial/pairwise distances values extracted by the Prony-based distance estimation method, and their distance distributions recovered by the distance distribution estimator method. Finally, based on the recovered distance distributions, the locations of the point sources are reconstructed. We show that the proposed approach is robust to additive white Gaussian noise for various simulation set-ups. 

%\clearpage
\newpage
\bibliographystyle{IEEEbib}
\bibliography{references.bib}

\end{document}